\documentclass[12pt]{article}
\usepackage{graphicx}
\usepackage{dcolumn}
\usepackage{bm}
\usepackage{subfigure}
\usepackage{color}
\usepackage{amsmath}
\usepackage{amssymb}
\usepackage{multirow}
\usepackage{scicite}
\usepackage{times}

\newenvironment{sciabstract}{%
\begin{quote} \bf}
{\end{quote}}

\title{Femtoscale magnetically induced lattice distortions in multiferroic TbMnO$_3$}

\author{H. C. Walker$^1$, F. Fabrizi$^{1,2,3}$, L. Paolasini$^1$, F. de Bergevin$^1$, J. Herrero-Martin$^1$\\
A. T. Boothroyd$^3$, D. Prabhakaran$^3$ \& D. F. McMorrow$^2$\\
\normalsize{$^{1}$European Synchrotron Radiation Facility, B\^{o}ite Postale 220, 38043 Grenoble, France}\\
\normalsize{$^{2}$London Centre for Nanotechnology and Department of Physics and Astronomy, }\\
\normalsize{University College London, Gower Street, London WC1E 6BT, UK}\\
\normalsize{$^{3}$Department  of Physics, Clarendon Laboratory, University of Oxford, Oxford, OX1 3PU, UK}}

\date{}

\begin{document}

\baselineskip24pt

\maketitle

\begin{sciabstract}
Magneto-electric multiferroics exemplified by TbMnO$_3$ possess both magnetic and ferroelectric long-range order.  The magnetic order is mostly understood, whereas the nature of the ferroelectricity has remained more elusive.  Competing models proposed to explain the ferroelectricity are associated respectively with charge transfer and ionic displacements. Exploiting the magneto-electric coupling, we use an electric field to produce a single magnetic domain state, and a magnetic field to induce ionic displacements.  Under these conditions, interference charge-magnetic X-ray scattering arises, encoding the amplitude and phase of the displacements. When combined with a theoretical analysis, our data allow us to resolve the ionic displacements at the femtoscale, and show that such displacements make a significant contribution to the zero-field ferroelectric moment.

\end{sciabstract}

The discovery of spin-cycloid multiferroics, in which the onset of non-collinear magnetic order
leads to a spontaneous ferroelectric polarization, has generated considerable interest in the control of electric polarization by magnetic fields, and vice versa\cite{KimuraTMO,Cheong2007}.  While comprehensive, microscopic
descriptions of their magnetic structures have been obtained\cite{kenzelmann,Arima,Fabrizi,AliouanePRL,Wilkins,Lawes,Prokhnenko,Feyerherm,harris}, our understanding of the ferroelectric state is still emerging. Two competing theoretical scenarios have been proposed: one purely electronic, without ionic displacements\cite{Katsura2005}; one based on anti-symmetric exchange interactions, with ionic displacements\cite{Sergienko}. Experiments have been unable to resolve the individual ionic displacements\cite{Bridges}.

Spin-cycloid multiferroics exhibit an exceptionally strong cross-coupling between the different types of order, as demonstrated when the electric (magnetic) field $\mathbf{E}$ ($\mathbf{H}$) was used to control magnetization $\mathbf{M}$ (ferroelectric polarization $\mathbf{P}$)\cite{KimuraTMO,KimuraPRB,yamasaki,Cabrera,Fabrizi,nivo}. Interest in this class of multiferroic has been generated both by the potential for novel devices, and the challenge they represent to our fundamental understanding of ordering phenomena in solids.
TbMnO$_3$ is the prototypical spin-cycloid multiferroic\cite{KimuraTMO}. Diffraction studies have established that in its ferroelectric phase below
$\sim$30~K the Mn magnetic moments form a cycloid in the \emph{bc} plane\cite{kenzelmann}, while the Tb moments order sinusoidally \cite{Fabrizi}, Fig.\ 1A,B. Formation of the cycloid removes the centre of inversion at the Mn sites and generates a spontaneous $\mathbf P$
along $c$. The scenario in which $\mathbf P$ is generated by ionic displacements has been investigated by ab-initio density functional theory (DFT)\cite{Xiang,Malashevich} which makes definite predictions for the displacements of the constituent ions.  Experimentally, only an upper limit (of $\sim$500~fm) has been estimated for the ionic displacements from EXAFS measurements\cite{Bridges}; EXAFS has been used to obtain femtoscale displacements in other systems\cite{Pettifer}. Application of a sufficiently strong magnetic field along either the $a$ or $b$ axis results in the flopping of   $\mathbf P$ from the $c$ to $a$ axis\cite{KimuraPRB,AliouanePRL}. Conventional X-ray scattering in an applied magnetic field has revealed charge reflections at both the first and second harmonics\cite{Aliouane06,Strempfer}.

Figure\ 1C--E provides a schematic of the X-ray diffraction technique we developed to  determine ionic displacements with improved accuracy.
When interacting with a solid, X-rays may be scattered by both the charges and any magnetic moments. Sufficiently far from atomic absorption edges the observed intensity $I$ may be written in the simplified form
\begin{equation}
\label{eq1}
I\propto\left|F_{\delta}-i\left(\frac{\cal E}{mc^2}\right)F_M\right|^2,
\end{equation}
where $\cal E$ is the X-ray energy, and $mc^2$ is the electron rest mass energy\cite{note2}.
Here $F_M$ is the non-resonant magnetic scattering (NRMS) amplitude\cite{Fabrizi}, and $F_{\delta}\propto\sum_jf_j(\mathrm{Q},{\cal E})\exp(i\mathbf{Q}\cdot[\mathbf{r}_j+\delta\mathbf{r}_j])$ is the scattering amplitude arising from any lattice
distortion ($\delta\mathbf{r}_j$), with $\mathbf Q$ and $\mathbf r_j$ being the wave-vector transfer and atomic position of the $j$'th atom, respectively\cite{Lovesey}.
(Note that $F_{\delta}=0$ for $\delta\mathbf{r}_j=0$, $\mathbf{Q}$ incommensurate).
The atomic form factor $f_j(\mathrm{Q},{\cal E})=f^0(\mathrm{Q})+f^\prime({\cal E})+if^{\prime\prime}({\cal E})$ depends on $\mathrm{Q}$ through the Thomson term $f^0(\mathrm{Q})$ and on ${\cal E}$
through the dispersion corrections $f^\prime({\cal E})$ and $f^{\prime\prime}({\cal E})$. Figure\ 1C refers to the case of
charge scattering, where no additional information is gained by reversing the handedness of the incident X-rays.
However, for the case of scattering from the magnetic cycloid in TbMnO$_3$ there is a symmetric reversal
in the intensities for left circular polarised (LCP) and right circular polarised (RCP) X-rays (Fig.\ 1D), allowing the population of right and left handed cycloidal
domains to be determined\cite{Fabrizi,nivo}. When an applied
magnetic field induces an atomic displacement with the same modulation period as the cycloid (Fig.\ 1E), the charge and magnetic amplitudes in Eq.\ 1 interfere  leading to a distinctive change in the polarization state of the scattered beam, reflected by changes in the Stokes parameters.

In zero applied magnetic field, the sample was cooled in an electric field to $T=15$~K within the ferroelectric phase. The intensity $I(\eta)$ was measured at the reciprocal lattice location ($h$,$k$,$l$)=(4,$\tau$,$-$1), where $\tau$ is the modulation wavevector of the magnetic structure (Fig.\ 1A,B), and ($h$,$k$,$l$) are the Miller indices.
This established that the electric field had produced an imbalance in the populations of the two cycloidal domains ($92\%:8\%$)\cite{Fabrizi,nivo}. A magnetic field was then applied along the $b$ axis and the (4,$\tau$,-1) reflection tracked for both LCP and RCP X-rays incident on the sample.

Figure\ 2A illustrates that when applying a magnetic field the scattering process in the sample changes to produce a scattered beam that is more linearly polarized, as reflected by the increase in $P_1$ stated for fits to the RCP data.
Attempts to model $I(\eta)$ using calculations based on distortions of the zero-field magnetic structure were unsuccessful. Instead, we consider that the magnetic field induces an additional uniform magnetic moment on the Tb and Mn ions polarized along the $b$ axis. In the presence of magneto-elastic coupling this will in turn induce a lattice distortion with the same period, or modulation wavevector $\tau$, as the zero-field magnetic structure with an associated non-zero charge scattering amplitude $F_{\delta}$. The evolution of the intensities with magnetic field evident in
Fig.\ 2A therefore arises from an interference between $F_{\delta}$ and $F_M$ in Eq. 1.
The lines in Fig. 2A represent the result of fitting Eq. 1 for the real and imaginary components of $F_{\delta}$.
This approach provides a very satisfactory description of the data. The field dependence of $F_{\delta}$ extracted from the fits
(Fig. 2B) displays the linear dependence on applied magnetic field expected for magneto-elastically induced lattice displacements.
It is worth emphasizing that the scattering amplitude $F_\delta$ is measured through the interference with the known magnetic structure factor, rather than through an intensity, and is therefore obtained with its phase, and is normalized in absolute units.

In order to identify which particular field-induced atomic displacement modes give rise to the charge scattering, $I(\eta)$ was measured for the two magnetic reflections (4,$\tau$,1) and (4,$\tau$,$-$1), at three different X-ray energies of $6.16$, $6.85$ and $7.77$~keV. This enabled us to exploit
the significant variation of the atomic dispersion corrections for TbMnO$_3$ (Fig. 3A). The lines in Fig. 3B represent fits to Eq. 1 allowing a free variation of the complex variable $F_{\delta}$. An important feature of the data is that the charge scattering is not invariant under inversion of the sign of the Miller index $l$ of the reflections.

The physical origin of the charge scattering measured in our experiments is magnetostrictive atomic displacements resulting from interactions between the zero field moments ($\mathbf{m(x)}$) and the uniform field induced moments ($\Delta\mathbf{m}$) \cite{note2}. Given the symmetry of the magnetic structure, which splits into two parts belonging to different irreducible representations (irreps) $\Gamma_3$ and $\Gamma_2$, and the fact that the induced magnetic moment $\Delta \mathbf{m}$  for a field along $b$ transforms as $\Gamma_2$ \cite{note2}, the modulated displacement modes also belong to two classes, described by the $\Gamma_4(=\Gamma_3\otimes\Gamma_2)$ and $\Gamma_1=(\Gamma_2\otimes\Gamma_2)$ irreps. These classes arise from interactions of $\Delta\mathbf{m}$ with the Mn moment component along y, and from interactions with all the other zero field moment components, respectively \cite{note2}.

Analysis of the displacement structure factor shows that it is even with respect to the signs $h=\pm4$ and $k=\pm\tau$, and is also even for $l=\pm1$ for the $\Gamma_4$ part, but odd for the $\Gamma_1$ part\cite{note2}. Therefore, by adding and subtracting the structure factors measured for $l=\pm1$, one can separate out the two components. The $\Gamma_1$ part of the geometrical structure factor is purely imaginary whereas the $\Gamma_4$ one is purely real.
Having a purely real or imaginary geometrical structure factor allows one to directly attribute the real and imaginary parts of the $F_\delta$ derived from
fits to the data to the corresponding parts of the atomic scattering factors. The  drastic variation with the energy of $f''$ for Mn and Tb (Fig. 3A), enables us to separate the contributions of the different atoms.

Taking the fitted values of the complex charge scattering amplitudes extracted from the data, we obtain three sets of simultaneous equation at the three different energies for each
of the irreps. Solving these equations we obtain values of the
amplitude and phase of the ionic displacements \cite{note2}.  For the $\Gamma_4$ irrep we find that the
displacements associated with the oxygen ions are dominant.
Both  inequivalent oxygen sites (O1 and O2) can in principle contribute within the constraint
$\delta_c^{O1}-10\delta_{a}^{O2}\approx-72\pm4$ in units of femtometres per Tesla (fm/T),
where for simplicity we have retained only the most significant contributions.
Within the same irrep,
the displacements of the other ions are given respectively by
$\delta_c^{Tb}$=$-19\pm2$ and $i\delta_b^{Mn}$=$5\pm5$ in fm/T. The factor of $i$ in the
Mn displacement indicates that it occurs in quadrature with the phase referenced with respect
to the maximum of the  $b$ component of the zero-field Mn magnetization (Fig.\ 4A). For the
$\Gamma_1$ irrep, we also find that the displacement of the oxygen ions dominate,
constrained by $\delta_c^{O2}-0.34i\delta_{b}^{O1}\approx-50\pm20$, with those of
the remaining ions given by $i\delta_b^{Tb}\approx-6\pm6$, and
$\delta_c^{Mn}\approx-4\pm4$, again in units of fm/T.

The interaction between the field-induced magnetic moments and the zero-field magnetic structure can be mediated via both symmetric and antisymmetric interactions, i.e. $\boldsymbol\delta\propto\Delta\mathbf m\cdot\mathbf m (\mathbf x)$, or $\boldsymbol\delta\propto\Delta\mathbf m\times\mathbf m (\mathbf x)$. We find that the $\Gamma_4$ displacements arise from a symmetric interaction between the $b$ components of both $\Delta\mathbf{m}$ and the Mn cycloid, whereas the $\Gamma_1$ displacements arise from both symmetric and anti-symmetric Dzyaloshiskii-Moriya type interactions between $\Delta\mathbf{m}$ and the other pre-existing $a$, $b$, $c$ components of $\mathbf{m}$, rendering their analysis more complex.

In order to compare our results with earlier work we now consider the $\Gamma_4$ Tb ($\delta^{Tb}_c$) displacements, illustrated in Fig.\ 4B (right panel).
Arising from a symmetric interaction, these displacements of the Tb ions are in anti-phase along $c$, and hence this particular displacement mode does not produce any additional ferroelectric
polarization, consistent with the plateau seen in measurements of the electric polarization as a function of magnetic field below the critical transition field\cite{KimuraTMO}.

Our results also shed light on the ionic displacements present in the zero-field ferroelectric phase (Fig. 4B, left panel)\cite{note2}.
In zero-field the Tb moments along $b$ are staggered with an amplitude of $\sim 1.0\mu_B$\cite{Fabrizi}. From magnetization measurements\cite{KimuraTMO}, we estimate the field required to produce the same amplitude of moment to be $H=1.1$ T, and thus the zero-field average ionic displacement of the Tb ions along  $c$  to be $1.1\times -19$=$-21\pm3$ fm.  By virtue of the symmetric nature of the interaction, the Tb displacements in zero-field are in phase along the $c$ axis, and hence contribute to the macroscopic polarization (Fig.\ 4B, left panel). We note that our estimate of the magnitude of the zero-field Tb displacement is similar to that derived from DFT calculations\cite{Malashevich}. Assuming a Tb charge of +3 and a displacement of $-21\pm3$ fm,  the contribution to the electrical polarization from this mode is calculated to be  P=176~$\mu$Cm$^{-2}$, around a quarter of the measured value\cite{KimuraTMO} of P$\simeq600$~$\mu$Cm$^{-2}$. This establishes that ionic displacements of the symmetry and magnitude determined in our study account for the spontaneous ferroelectric polarization in TbMnO$_3$ to within an order of magnitude. Our data decisively support microscopic models which attribute $\mathbf P$ to ionic displacements, but they also point to the need to  include both symmetric and anti-symmetric magnetic interactions in any such models. The technique introduced here should be applicable to other multiferroics, and to the wider class of systems with complex order in the presence of magnetoelastic coupling.

\textbf{Acknowledgements}: We thank Andrea Fondacaro, Claudio Mazzoli and Gilbert Pepellin for their experimental assistance and Yves Joly, Andrei Rogalev and Christian Vettier for fruitful discussions. We also thank EPSRC and the Royal Society for financial support. The data are stored at the European Synchrotron Radiation Facility.

\clearpage

\begin{figure}
\centering
\includegraphics[width=0.9\textwidth]{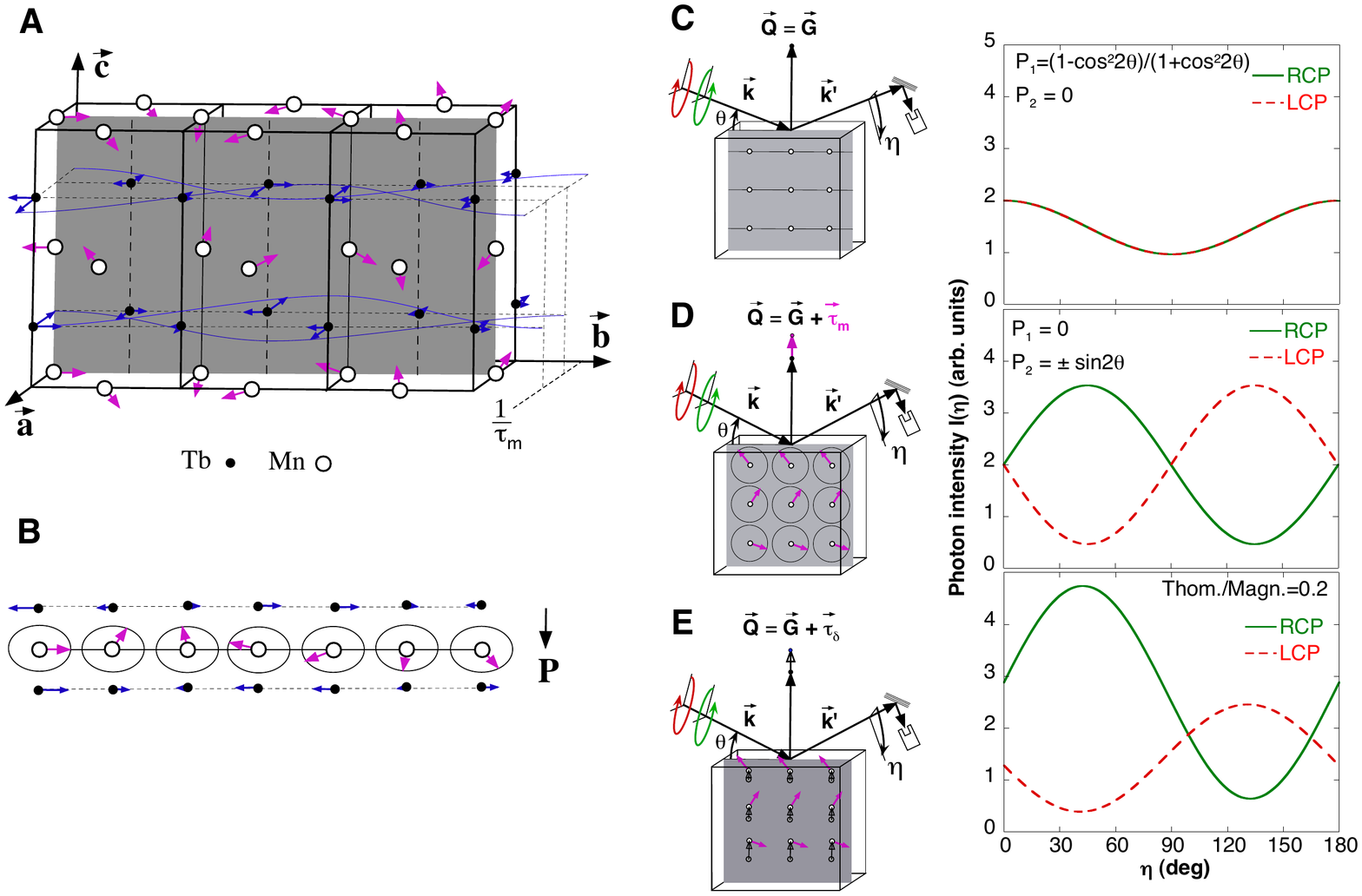}
\caption{{\bfseries The X-ray diffraction experiment}
(A) The crystallographic and magnetic structure of TbMnO$_3$, where arrows in blue correspond to the spin moments on the Tb, and the arrows in pink to the those on the Mn. (B) Projecting the magnetic structure onto the bc plane, shows how the Mn spin cycloid relates to the spontaneous ferroelectric polarization $\mathbf{P}$. (C--E) Incident circularly polarized X-rays diffract from a sample and the polarization state of the scattered beam is determined using a linear polarization analyser.
$\mathbf k$ ($\mathbf k^\prime$) is the wavevector of the incident (scattered) beam.
The intensity $I(\eta)$ as a function of the rotation angle $\eta$ around $\mathbf k^\prime$
is given by $I(\eta)\propto1+P_1\cos(2\eta)+P_2\sin(2\eta)$, where the Stokes parameter $P_1$ describes the linear polarization $\|$ and $\bot$ to the plane of scattering, and $P_2$, the oblique linear polarization. (C) Charge diffraction occurs when the wavevector transfer $\mathbf Q=\mathbf k^\prime-\mathbf k$
is equal to a reciprocal lattice vector $\mathbf G$ (Laue condition). (D) Non-resonant magnetic scattering (NRMS) from an incommensurate cycloid
of propagation wavevector $\mathbf \tau_m$ depicted with pink arrows. (E) Interference between NRMS and charge scattering ($20\%$ that of NRMS) due to atomic displacements shown with black arrows ($\tau_{\delta}=\tau_m$).}
\end{figure}

\begin{figure}
\centering
\includegraphics[width=0.9\textwidth]{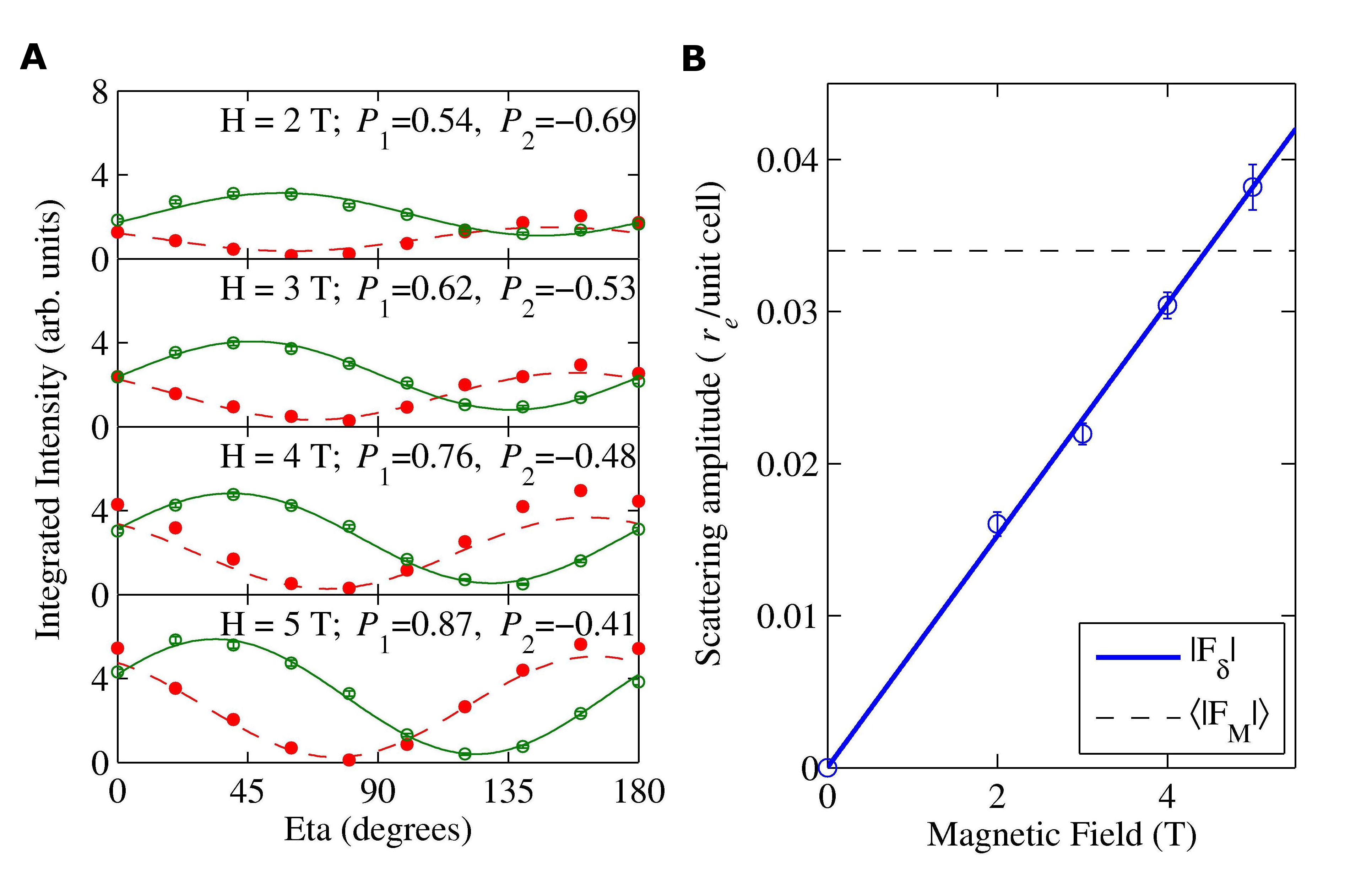}
\caption{{\bfseries Field and polarization dependence of the interference scattering}
(A) $I(\eta)$ at (4,$\tau$,-1) of TbMnO$_3$ at $T=15$~K for ${\cal E}=6.85$~keV, for different magnetic fields applied along the $b$ axis, with LCP (solid red circles) and RCP (open green circles) incident. The lines are fits to Eq. 1 from which the charge scattering amplitude $F_{\delta}$ can be extracted. (B) $|F_{\delta}|$ in units of the classical electron radius $r_e$ as a function of applied magnetic field. The dashed line corresponds to the NRMS amplitude, $\langle |F_{M}|\rangle$, averaged over the final polarization states of the scattered beam.}
\end{figure}

\begin{figure}
\centering
\includegraphics[width=0.9\textwidth]{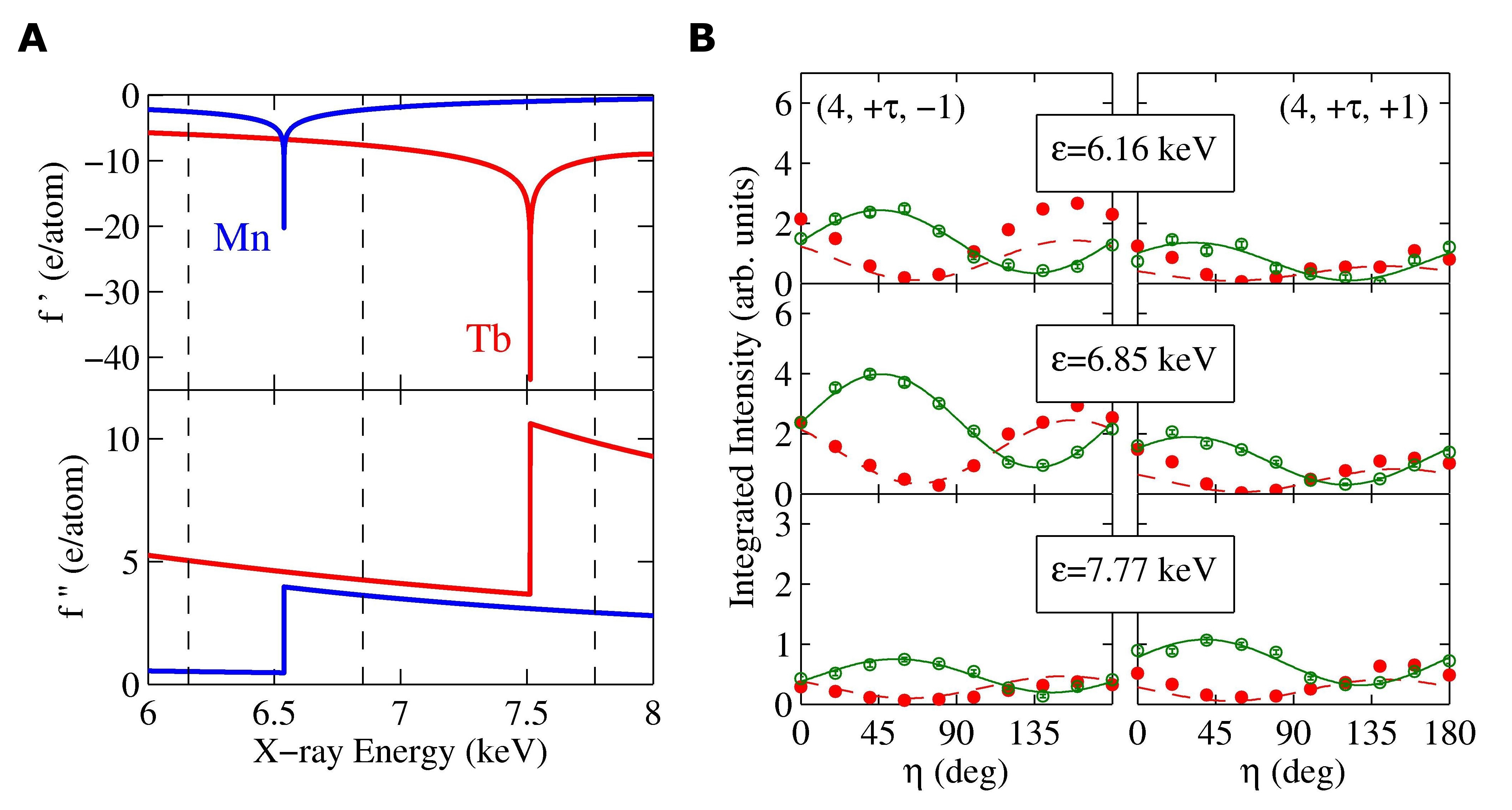}
\caption{{\bfseries Dependence of the scattering on X-ray energy} (A) The calculated dispersion corrections ($f^\prime$, $f^{\prime\prime}$) for Tb and Mn as a function of energy. (B) $I(\eta)$ for (4,$\tau$,$\pm$1) at $H//b=3$~T, for different energies with LCP (solid red circles) and RCP (open green circles) X-rays incident. The lines are fits to Eq. 1 computed separately for each reflection at each X-ray energy.}
\end{figure}

\begin{figure}
\centering
\includegraphics[width=0.9\textwidth]{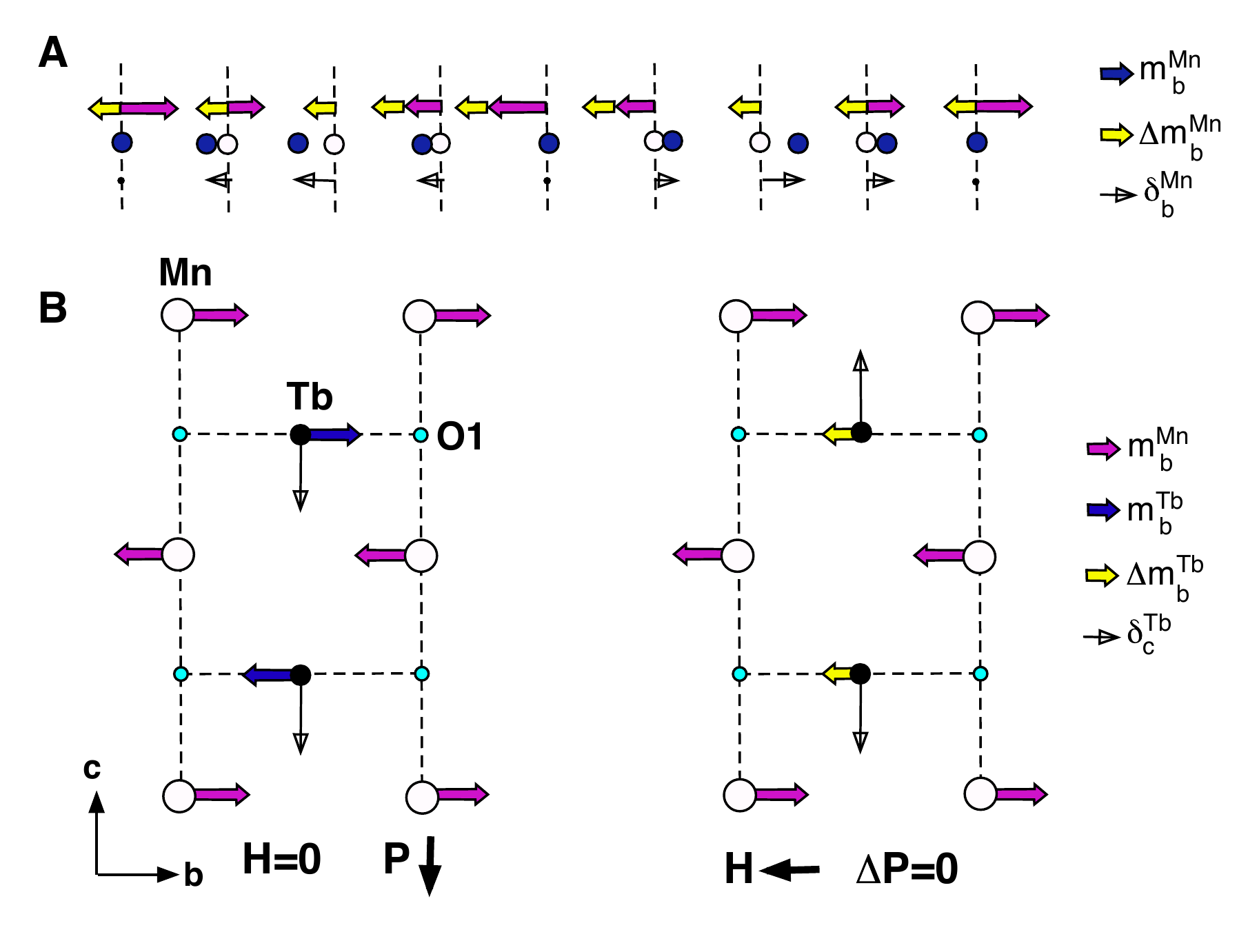}
\caption{{\bfseries Ion displacements in TbMnO$_3$}
(A) Mn displacements ($\delta^{Mn}$) vary along $b$ because of an imbalance in neighbouring magnetization densities resulting from the combination of the projection of the Mn spin cycloid on the $b$ axis ($m^{Mn}_b$) and the uniform induced moment ($\Delta m^{Mn}_b$). Note that $\delta^\mathrm{Mn}_b$ is in quadrature with the magnetization.
(B) Magnetoelastic distortions of the Tb atoms ($\delta^{Tb}_c$) along the $c$ axis in zero magnetic field (left) and for a field applied along the $b$ axis (right). In zero field the interaction between the b-components of the Tb moment ($m^{Tb}_b$) and of the Mn moment ($m^{Mn}_b$) via the oxygen atoms results in the distortion $\delta^{Tb}_c(0)$ producing a non-zero spontaneous electric polarization along the $c$ axis. For field applied along the $b$ axis, a uniform moment is induced on the Tb ($\Delta m^{Tb}_b$). The interaction between this and the Mn moment ($m^{Mn}_b$) results in the distortion $\delta^{Tb}_c(H)$, which alternates on moving along the $c$ axis, leading to zero additional electrical polarization.}
\end{figure}

\clearpage

\setcounter{figure}{0}
\makeatletter
\renewcommand{\thefigure}{S\@arabic\c@figure}

\makeatletter
\renewcommand{\thetable}{S\@arabic\c@table}

\setcounter{equation}{0}
\makeatletter 
\def\tagform@#1{\maketag@@@{(S\ignorespaces#1\unskip\@@italiccorr)}}
\makeatother

\section*{Supporting Online Material for: Femtoscale magnetically induced lattice distortions in multiferroic TbMnO$_3$}

\subsection*{Methods}

Experiments were performed at the ID20 Magnetic Scattering Beamline ({\it 26}) at the European Synchrotron Radiation Facility using a single crystal of TbMnO$_3$ ({\it Pbnm} space group $\sharp62$ $a=5.315$~\AA, $b=5.831$~\AA, $c=7.377$~\AA) synthesized at the University of Oxford using the floating zone method. The sample was glued using conductive silver paint between two copper electrodes to allow an electric field to be applied to the sample of up to $4$~kV/mm. The sample stick was then inserted into an Oxford Instruments $10$~T cryomagnet. The mounting of the sample was such as to have $[100]$ direction specular, with an electric field applied along $[00\bar{1}]$ (the sense of the electric field and the polarization shown in figures in Fabrizi et al. ({\it 5}) is incorrect and should be reversed, see ({\it 27})) and the magnetic field applied along $[010]$, resulting in a horizontal scattering place described by $[100]-[001]$. A voltage of $545$~V was applied across the sample as it was cooled from $T=60$~K to $15$~K.

The incident horizontally linearly polarized beam was converted into a circular polarization state using an in-vacuum quarter-wave $720$~$\mu$m diamond phase plate ({\it 28}). Despite the use of different incident energies, this thickness of diamond allowed us to produce a beam which was $99$~\% circularly polarized. The polarization state of the scattered beam was obtained using single crystal analyzers as appropriate for the different incident energies (LiF(220) at $E=6.16$~keV, Cu(220) at $E=6.85$~keV and Pt(222) at $E=7.77$~keV). The background signal was measured after deplacing the sample by $\Delta\theta=2$ degrees, enabling it to be subtracted from the total signal.

On application of the magnetic field the wave-vector $\tau$ was found to remain constant below $7$~T, before starting to reduce in magnitude, implying the onset of the polarization-flop transition ({\it 15}). Also below the critical field no major change of the NRMS for $\eta=90^\circ$ was observed, indicating that the magnetic field does not perturb appreciably the zero-field domain populations.

For both symmetry classes of ionic displacements, the greatest distortions are those associated with the oxygens. However, our current measurements provide values for the combined oxygen displacements, although we note that in principle this ambiguity could be removed by measuring additional reflections.

\subsection*{Analysis}

\subsubsection*{Symmetry Analysis}

The application of the magnetic field along the $b$ axis induces a uniform moment $\Delta\mathbf{m}^{Mn(Tb)}$ of the Mn (Tb) lattice, ignoring any modulated moments possibly also induced. Due to exchange striction or magnetocrystalline anisotropy, $\Delta\mathbf{m}^{Mn(Tb)}$ in combination with the zero field magnetic moments, $\mathbf{m(x_1)}$ at $\mathbf{x_1}$, produces atomic displacements, where to lowest order the displacement of an atom at $\mathbf{x_2}$ is
\begin{equation}\label{eqS1}
\boldsymbol\delta\mathbf{(x_2)}=\sum_\mathbf{x_1}\mathbf{m(x_1)}\left(\xi_{Mn}(\mathbf{x_2},\mathbf{x_1})\Delta\mathbf{m}^{Mn} + \xi_{Tb}(\mathbf{x_2},\mathbf{x_1})\Delta\mathbf{m}^{Tb}\right).
\end{equation}
The tensor $\xi(\mathbf{x_2},\mathbf{x_1})$ combines the coefficients of two free energy terms, a term of magnetostriction $\boldsymbol\delta\mathbf{(x_2)m(x_1)}\Delta\mathbf{m}$, and a term of stiffness $\boldsymbol\delta^2\mathbf{(x_2)}$, and is invariant under the symmetry elements of the space group {\it Pbnm}. $\mathbf{m(x_1)}$ is comprised of parts $\mathbf{m}_2$ and $\mathbf{m}_3$ belonging to two different irreducible representations (irreps). $\mathbf{m}_3$ describes the manganese component $m_b^{Mn}$ which belongs to the irrep ($\tau\mathbf{b}$,$\Gamma_3$), whilst $m_c^{Mn}$, $m_a^{Tb}$, and $m_b^{Tb}$ (as first determined by non-resonant X-ray scattering ({\it 5})), part $\mathbf{m}_2$, belong to ($\tau\mathbf{b}$,$\Gamma_2$), where $\tau\mathbf{b}$ is the propagation vector and $\Gamma_{3(2)}$ the little representation ({\it 3, 11, 29, 30}). $\Delta\mathbf{m}$ on the other hand is an axial vector along $b$, and hence it changes sign under the action of mirrors $m_{xy}$ and $m_{yz}$, and remains unchanged under $2_y$, and so belongs to the little representation $\Gamma_2$ (see Table S1), but has a null propagation vector.
Therefore on inserting $\mathbf{m_3(x_1)}$ into the right hand of equation~\eqref{eqS1} we find that the displacement $\boldsymbol\delta\mathbf{(x_2)}$, induced by the symmetric interaction between $m_b^{Mn}$ and $\Delta\mathbf{m}$, belongs to the representation ($\tau\mathbf{b}$,$\Gamma_4$). Meanwhile, for $\mathbf{m_2(x_1)}$, symmetric and anti-symmetric interactions between $\Delta\mathbf{m}$ and the other $a$, $b$, and $c$ components of the zero magnetic field structure lead to displacements belonging to ($\tau\mathbf{b}$,$\Gamma_1$). Hence in both cases the displacements are visible at $\tau\mathbf{b}$ in reciprocal space.

\begin{figure}
\centering
\includegraphics[width=.8\textwidth]{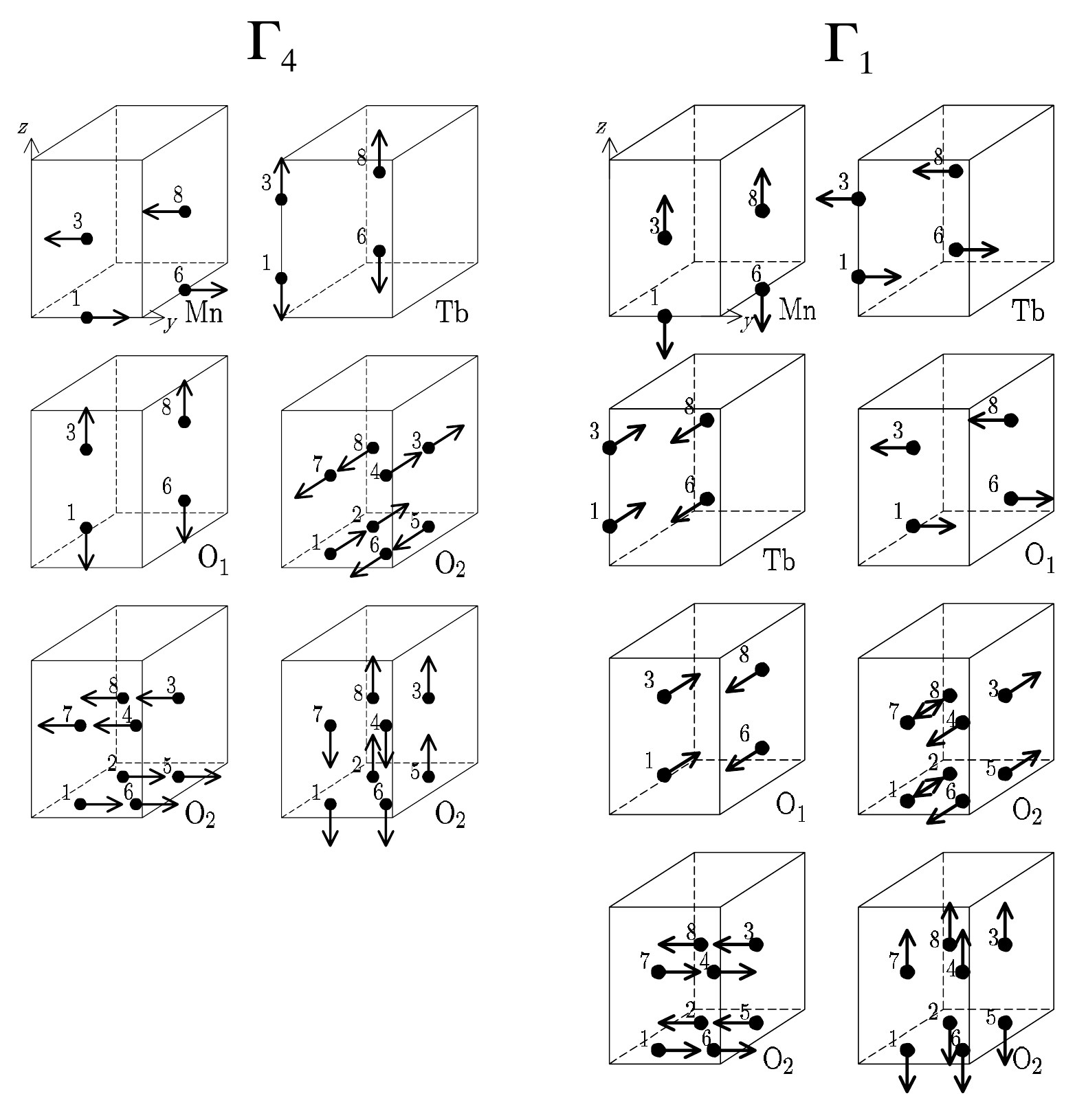}
\caption{Displacement modes in $\Gamma_1$ and $\Gamma_4$ visible at (4 $\tau$ $\pm1$).}\label{fig_modes}
\end{figure}

The space group {\it Pbnm} has eight atomic positions, but in the cycloidal ferroelectric phase the symmetry is reduced leading to the eightfold position being split into two independent fourfold orbits. The individual atomic displacements can then usefully be combined into two sets of modes:
\begin{align}
\mathbf{\Delta}_{\alpha1}&=\boldsymbol\delta_1+\boldsymbol\delta_3+\boldsymbol\delta_6+\boldsymbol\delta_8&\mathbf{\Delta}_{\alpha2}&=\boldsymbol\delta_5+\boldsymbol\delta_7+\boldsymbol\delta_2+\boldsymbol\delta_4\nonumber\\
\mathbf{\Delta}_{\beta1}&=\boldsymbol\delta_1-\boldsymbol\delta_3+\boldsymbol\delta_6-\boldsymbol\delta_8&\mathbf{\Delta}_{\beta2}&=\boldsymbol\delta_5-\boldsymbol\delta_7+\boldsymbol\delta_2-\boldsymbol\delta_4\nonumber\\
\mathbf{\Delta}_{\gamma1}&=\boldsymbol\delta_1+\boldsymbol\delta_3-\boldsymbol\delta_6-\boldsymbol\delta_8&\mathbf{\Delta}_{\gamma2}&=\boldsymbol\delta_5+\boldsymbol\delta_7-\boldsymbol\delta_2-\boldsymbol\delta_4\nonumber\\
\mathbf{\Delta}_{\delta1}&=\boldsymbol\delta_1-\boldsymbol\delta_3-\boldsymbol\delta_6+\boldsymbol\delta_8&\mathbf{\Delta}_{\delta2}&=\boldsymbol\delta_5-\boldsymbol\delta_7-\boldsymbol\delta_2+\boldsymbol\delta_4,
\end{align}
where the numbering of the atomic positions follows the International Table of Crystallography notation for spacegroup written in the {\it Pbnm} setting, and which can also be written according to:
\begin{equation}
\mathbf{\Delta}_{\alpha\pm}=\frac{1}{2}(\mathbf{\Delta}_{\alpha1}\pm\mathbf{\Delta}_{\alpha2}).
\end{equation}
These are the different modes listed in Table S2, showing the representation to which they belong, and the extinction rules for the reflections ($4$ $\pm\tau$ $\pm1$). The modes are not active for all the atoms, since Mn, Tb and O$_1$ are in special crystallographic positions. The active modes are depicted in Figure~\ref{fig_modes}.

\subsubsection*{Structure Factors}
If $A,j$ is the crystallographic position within the cell, where $A=M,T,P$ and $O$ label the atoms Mn, Tb, O$_1$ and O$_2$, and $j$ runs up to four or eight depending on the multiplicity of the site, such that $\mathbf{x}_{A,j}$ describes the initial position of the atom in the cell, whilst $\mathbf{u}$ is the position in the Bravais lattice, and $\mathbf{u}_o$ the origin of the lattice, then the absolute position of the atom is
\begin{equation}
\mathbf{x}+\boldsymbol\delta\mathbf{(x)}=\mathbf{u}_0+\mathbf{u}+\mathbf{x}_{A,j}+\boldsymbol\delta\mathbf{(x)},
\end{equation}
with its displacement
\begin{equation}
2\boldsymbol\delta\mathbf{(x)}=\boldsymbol\delta_{A,j}\exp(-2\pi i\mathbf{\tau\cdot x})+\boldsymbol\delta^*_{A,j}\exp(2\pi i\mathbf{\tau\cdot x}).
\end{equation}
The structure factor of the position $A$, with $\mathbf{K}=(40l)$, is
\begin{equation}
F_A(\mathbf{K}+\mathbf{\tau})=f_A\sum_{\mathbf{u},j}\exp\left(2\pi i(\mathbf{K}+\mathbf{\tau})\cdot(\mathbf{x}+\boldsymbol\delta\mathbf{(x)})\right).
\end{equation}
Since the crystal is composed of $N$ cells then to the lowest order approximation in $\boldsymbol\delta(\mathbf{x})$
\begin{align}
F_A(\mathbf{K}+\mathbf{\tau})&\approx Nf_AC_0(\pi i(\mathbf{K}+\mathbf{\tau}))\cdot\sum_j\boldsymbol\delta_{A,j}\exp2\pi i\mathbf{K\cdot x}_{A,j}\nonumber\\
F_A(\mathbf{K}-\mathbf{\tau})&\approx Nf_AC_0(\pi i(\mathbf{K}-\mathbf{\tau}))\cdot\sum_j\boldsymbol\delta^*_{A,j}\exp2\pi i\mathbf{K\cdot x}_{A,j}
\end{align}
where $C_0=\exp-2\pi i\mathbf{K\cdot u_0}$. It thus becomes clear that on changing the sign of $\tau$, $\boldsymbol\delta$ is replaced by it's complex conjugate. However, as shown in Table~S2, for the modes visible at (4 $\pm\tau$ $\pm1$) $\delta_{a,c}$ is real and $\delta_b$ is imaginary, and therefore the structure factor is invariant under this change of sign. We can now rewrite this structure factor for the {\it Pbnm} structure, where we separate out the parts associated with the $\Gamma_1$ and $\Gamma_4$ modes:
\begin{align}
F_{\Gamma_1}(4,\tau,\pm l) = &\pm N\pi(if_M\Delta_{M,\beta1,c}\nonumber\\
                             &-f_T(4i\sin8\pi x_T\Delta_{T,\gamma1,a}+\tau\cos8\pi x_T\Delta_{T,\beta1,b})\nonumber\\
                             &-f_O(4i\sin8\pi x_P\Delta_{P,\gamma1,a}+\tau\cos8\pi x_P\Delta_{P,\beta1,b})\nonumber\\
                             &-8if_O\sin8\pi x_O\sin2\pi z_O\Delta_{O,\gamma+,a}\nonumber\\
                             &-2\tau f_O\cos8\pi x_O\sin2\pi z_O\Delta_{O,\beta-,b}\nonumber\\
                             &+2if_O\cos8\pi x_O\cos2\pi z_O\Delta_{O,\beta+,c}),\label{Eq_G1}
\end{align}
\begin{align}
F_{\Gamma_4}(4,\tau,\pm l) = &N\pi(if_M\tau\Delta_{M,\beta1,b}-f_T\cos8\pi x_T\Delta_{T,\beta1,c}-f_O\cos8\pi x_P\Delta_{P,\beta1,c}\nonumber\\
                             &-8f_O\sin8\pi x_O\cos2\pi z_O\Delta_{O,\gamma-,a}\nonumber\\
                             &+2i\tau f_O\cos8\pi x_O\cos2\pi z_O\Delta_{O,\beta+,b}\nonumber\\
                             &-2f_O\cos8\pi x_O\sin2\pi z_O\Delta_{O,\beta-,c}),\label{Eq_G4}
\end{align}
where $\Delta_{M,\beta1}$ also reads as $\Delta_{M,\beta+}$, $\Delta_{T(P),\beta1}$ as $\Delta_{T(P),\beta-}$, and $\Delta_{T(P),\gamma1}$ as $\Delta_{T(P),\gamma+}$. Examination of equations~\eqref{Eq_G1} and \eqref{Eq_G4} reveals that the $\Gamma_1$ part of the geometrical structure factor is pure imaginary whilst the $\Gamma_4$ part is pure real, furthermore the symmetry imposes a phase on the modulated displacements, such that in the modes visible at ($4$ $\tau$ $\pm1$),
$b$ components are in quadrature with those along $a$ and $c$.

\subsubsection*{Extracting the magnetic field induced displacements}
To solve the simultaneous equations in the atomic displacements generated for the different energies a recursive procedure is used, considering the two classes separately. Since the real and imaginary parts of the structure factor are both known from the experiments, we have pairs of independent systems of equations containing separately the atomic factors $f^0+f^\prime$ on one side and $f^{\prime\prime}$ on the other. We first neglect the oxygen contributions $f^{\prime\prime}_O$ and solve the system for displacements of Mn and Tb only. We then turn to the system containing  $f^0+f^\prime$ and solve it for the oxygen displacements using the Tb and Mn positions just found. These two calculations steps are repeated, including now in the first step the oxygens at the positions previously obtained.
For a symmetric interaction between $m_b^{Mn}$ and $\Delta\mathbf{m}$, the resulting $\Gamma_4$ displacements in femtometers per Tesla are as follows:
\begin{align}
&\delta^\mathrm{Tb}_c = -19\pm2~\mathrm{fm/T},\nonumber\\
&i\delta^\mathrm{Mn}_b = 5\pm5~\mathrm{fm/T}\nonumber,\\
&\delta^\mathrm{O1}_c - 10*\delta^\mathrm{O2}_a - 0.37*i\delta^\mathrm{O2}_b - 0.36*\delta^\mathrm{O2}_c=-72\pm4~\mathrm{fm/T},
\end{align}
where $\delta^\mathrm{X}_j$ is the displacement of ion X along the $j$ axis. The reference for the phases of the displacements is defined by the maximum in the b-component of the Mn magnetization. $\delta_{a,c}$ are in phase or anti-phase with this, whilst $\delta_b$ are in quadrature, and are thus pure imaginary (see for example $\delta^\mathrm{Mn}_b$ in Fig. 4A). The  $\delta^{O2}_b$ displacement is likely to be small given that the exchange pathway is at $45^\circ$ as opposed to $90^\circ$. The displacements $\delta^{O2}_a$ and $\delta^{O2}_c$ would be entirely absent in the undistorted perovskite structure, and given the small coefficient of $\delta^{O2}_c$ these displacements can be assumed to be negligible, but the large coefficient of $\delta^{O2}_a$ is significant. Therefore we estimate that the displacements $\delta^{O1}_c$ and $10\times\delta^{O2}_a$ are dominant.

Now turning to the $\Gamma_1$ displacements arising from symmetric and antisymmetric interactions between the other zero field moment components and $\Delta\mathbf{m}$, solving the equations gives:
\begin{align}\label{eqS11}
&i\delta^\mathrm{Tb}_b + 6.8*\delta^\mathrm{Tb}_a = -6\pm6~\mathrm{fm/T},\nonumber\\
&\delta^\mathrm{Mn}_c = -4\pm4~\mathrm{fm/T},\nonumber\\
&\delta^\mathrm{O2}_c - 2.4*\delta^\mathrm{O1}_a - 0.34*i\delta^\mathrm{O1}_b + 3.2*\delta^\mathrm{O2}_a + 0.12*i\delta^\mathrm{O2}_b = -50\pm20~\mathrm{fm/T}.
\end{align}
This indicates that there are two different potential displacements of the Tb ions along $a$ and $b$, however in the undistorted perovskite structure the symmetry is such that the displacements along $a$ would be identically zero and therefore we assume that in the distorted structure these displacements will be smaller than those along $b$, and we neglect them. Then, for the five potential oxygen modes given in \eqref{eqS11}, by similar arguments we conclude that the dominant contributions are $\delta^{O2}_c$ and $\delta^{O1}_b$.

If we consider the magnetic field induced $\Gamma_4$ $\delta^{Tb}_c$ displacement, we find that it alternates along the $c$ axis, giving no additional ferroelectric polarization. However, in zero field the alternating direction of the Tb moments will result in displacements in phase along $c$, but to estimate the resultant ferroelectric polarization requires knowledge of the relative amplitudes of the field induced and zero field magnetic moments.

\subsubsection*{Converting displacements in field to those in zero field}
The symmetric interaction between the Tb and Mn moments is such that the displacement amplitude varies linearly as a function of the moment size. Starting from bulk measurements, it is shown that an applied field of $2$~T results in a magnetization of $1$~$\mu_B$/f.u. ({\it 1}).
Then to determine the moments induced on the Mn and Tb ions in the magnetic field, one needs to consider the magnetic quantum numbers $S$, $L$, and $J$. For Mn$^{3+}$: $S=2$, $L=0$ and $J=2$, while for Tb$^{3+}$: $S=3$, $L=3$ and $J=6$, allowing us to calculate the approximate Curie constant for each according to $C\propto g_J^2J(J+1)$, where $g_J$ is the Land\'{e} g-factor. The ratio of $C$ for Mn to Tb, taking into account that the value for Mn should be halved due to its modulated magnetization, estimates that the measured magnetization of $1$~$\mu_B$/f.u. equates to an induced moment of $\sim0.9\mu_B$ on the Tb in an applied field of $2$~T. This can then be compared to the magnitude of the moment component along $b$ for Tb in zero magnetic field: $\sim1.0\mu_B$ ({\it 5}). Therefore this suggests that the magnitude of the displacements in zero field will be of the same order as that extracted from our fits for an applied field of $2/0.9$~T. The extracted displacement is the maximum value, and moving along the $b$ axis the displacement will be modulated due to the modulation of the $b$ components of the Tb and Mn moments, so the average displacement will be the maximum divided by two.

\subsubsection*{Non-resonant and resonant magnetic scattering amplitudes}
In writing down Eq. 1 we state that this applies for photon energies sufficiently far from an
absorption edge such that resonant contributions to the X-ray magnetic scattering are negligibly small.
Here we establish the validity of this approximation by examining it from three different approaches.
First, we provide a theoretical estimate of the contributions from the dominant X-ray magnetic resonant scattering (XRMS)
channels. Second, we provide a brief review of relevant literature on experimental studies of resonant and non-resonant
X-ray magnetic scattering. Third, we provide a summary of a re-analysis of our data where we have included a significant
level of XRMS along with the non-resonant magnetic contribution.

First, we provide a theoretical estimate of the ratio of the resonant to the non-resonant X-ray magnetic scattering
lengths. In principle, we need to consider the resonant contributions from the Tb M and L edges, and the Mn L and K edges.
However, it is already established that the resonant enhancements of the scattering length at the
Tb L and Mn K edges are weak ({\it 31}). That leaves us with the Tb M$_4$ and M$_5$ ($\sim$1.2 keV) and the
Mn L$_2$ and L$_3$ ($\sim$0.6 keV) edges to consider for which the scattering length can be of order 100$r_e$, where $r_e$ is the classical radius of the electron ({\it 32}). The notion that the  contribution of XRMS from these edges may be significant, even in our experiments performed at energies roughly 5 keV away, follows from simple scattering theory which predicts that the contribution should fall off only slowly as the inverse of the difference in energy between the resonance and photon energies. Here we show that a full analysis of the theory of magnetic scattering reveals that the conclusion reached from the simple theory is wrong, and that the resonant scattering length falls off faster than predicted by simple theory. The full analysis is used to obtain estimates of the contributions from the Tb M$_{4,5}$ and Mn L$_{2,3}$ edges.

A consistent treatment of resonant and non resonant magnetic scattering has been
given by Blume ({\it 33}) (see also ({\it 34--36}).
In units of $r_e$, the scattering amplitude can be written as
\begin{eqnarray}
f(\hbar\omega) &=&
- \sum_i\{\langle a|\, e^{2 \pi i {\bf Q \cdot r}_i} |\,a\rangle
\boldsymbol{\epsilon'^*} \cdot \boldsymbol{\epsilon}
- i \frac{\hbar \omega}{mc^2}
\langle a|\, {\bf s}_i e^{2 \pi i {\bf Q \cdot r}_i} |\,a\rangle {\bf\cdot}
\boldsymbol{\epsilon'^*} \times \boldsymbol{\epsilon}\ \} \nonumber \\
&+& \frac 1m \sum_c \{ \frac{\boldsymbol{\epsilon'^*} \cdot
{\langle a|\, {\bf O}^\dag ({\bf k'})|\,c\rangle}
\langle c|\, {\bf O} ({\bf k})|\,a\rangle \cdot
\boldsymbol{\epsilon}}
{E_a - E_c + \hbar\omega - i \Gamma_c / 2} \nonumber \\
&+& \frac{\boldsymbol {\epsilon}\cdot
{\langle a|\, {\bf O} ({\bf k})|\,c\rangle}
\langle c|\,{\bf O}^\dag ({\bf k'}) |\,a\rangle \cdot
\boldsymbol{\epsilon'^*}}{E_a - E_c - \hbar\omega }\}
\label{App:Eq:scatt1}
\end{eqnarray}
Here $\bf k, \boldsymbol{\epsilon}, \hbar\omega$
are the wavevector, polarisation and energy of the incident photon;
the primed quantities refer to the scattered photon.
$|\,a\rangle$ is the ground state of the scattering object. The summations
are over all of its electrons $i$ and all of its excited states $|\,c\rangle$ of inverse
lifetime $\Gamma_c$. ${\bf Q} = {\bf k} - {\bf k'}$ is the wavevector transfer. The operators $\bf O$ are
\begin{equation}
{\bf O}({\bf k}) = \sum_i\ e^{ 2 \pi i {\bf k \cdot r}_i}
({\bf p}_i - i h {\bf k} \times {\bf s}_i)
\end{equation}

The first term in Eq.~\eqref{App:Eq:scatt1} is Thomson scattering, the second
one is a piece of the non-resonant magnetic scattering and the two last the
dispersive terms, one of them being resonant. Another part of the non-resonant
magnetic scattering is hidden in these resonant or dispersive terms
from which it must be extracted. After some algebra the scattering amplitude becomes
\begin{eqnarray}
f(\hbar\omega) &=&
- \sum_i\{\langle a|\, e^{2 \pi i {\bf Q \cdot r}_i} |\,a\rangle
\boldsymbol{\epsilon'^*} \cdot \boldsymbol{\epsilon} \nonumber \\
&-& i \frac{\hbar \omega}{mc^2}\, \boldsymbol{\epsilon'^*} \cdot
\langle a|\, e^{2 \pi i {\bf Q \cdot r}_i}
(-i \frac{{\bf Q} \times{\bf p}_i}{h {\bf Q}^2} \cdot {\bf A}
+ {\bf s}_i \cdot {\bf B} )|\,a\rangle \cdot
\boldsymbol{\epsilon}\ \} \nonumber \\
&-& \sum_c \frac {E_a - E_c}{m \hbar \omega} \{ \frac{\boldsymbol{\epsilon'^*}
\cdot {\langle a|\, {\bf O}^\dag ({\bf k'})|\,c\rangle}
\langle c|\, {\bf O} ({\bf k})|\,a\rangle \cdot
\boldsymbol{\epsilon}}
{E_a - E_c + \hbar\omega - i \Gamma_c / 2} \nonumber \\
&-& \frac{\boldsymbol {\epsilon}\cdot
{\langle a|\, {\bf O} ({\bf k})|\,c\rangle}
\langle c|\,{\bf O}^\dag ({\bf k'}) |\,a\rangle \cdot
\boldsymbol{\epsilon'^*}}{E_a - E_c - \hbar\omega }\}
\label{App:Eq:scatt2}
\end{eqnarray}
The second term is now the full non-resonant magnetic scattering amplitude, while
the last two terms represent the remaining dispersive amplitude.
They differ from Eq.~\eqref{App:Eq:scatt1}
by a factor $(E_a - E_c)/{\hbar \omega}$ and a change of sign of the first term.
Their numerators, as far as the operators $\bf O$ and polarisations
$\boldsymbol \epsilon$'s are concerned, are symmetric to each other by time
inversion. This results in their sum representing the non-magnetic properties
of the states $|\,a\rangle$ and $|\,c\rangle$, while their difference
represents the magnetic properties. To separate out the magnetic and
non-magnetic resonant scattering, we have to isolate the sum and difference.
This is given is Eq.\ (A8) of Ref. ({\it 33}) where the resonant
magnetic scattering amplitude appears as
\begin{eqnarray}
f_{res,\,mag}(\hbar\omega) =
- \sum_c \frac {(E_a - E_c)^2} {m \hbar \omega
((E_a - E_c)^2 - (\hbar\omega - i \Gamma_c / 2)^2)} \nonumber \\
\{\boldsymbol{\epsilon'^*}
\cdot {\langle a|\, {\bf O}^\dag ({\bf k'})|\,c\rangle}
\langle c|\, {\bf O} ({\bf k})|\,a\rangle \cdot
\boldsymbol{\epsilon}
- {\boldsymbol {\epsilon}\cdot
{\langle a|\, {\bf O} ({\bf k})|\,c\rangle}
\langle c|\,{\bf O}^\dag ({\bf k'}) |\,a\rangle \cdot
\boldsymbol{\epsilon'^*}}
\}
\label{App:Eq:resmagscat}
\end{eqnarray}
A minor approximation is based on $\Gamma_c/2 \ll |E_a - E_c - \hbar\omega|$.
The decrease of the tail at high energy is now much faster than for the main
resonant term in Eq.~\eqref{App:Eq:scatt1}. The physical reason for such a
behaviour is that all resonant tails have been depleted to build up the
non-resonant magnetic scattering. The amplitude which was expected from these
tails is inside the non-resonant term. Some doubt has also been
cast on the validity of the non-resonant magnetic scattering formula when the
radiation energy is not much larger than any resonance energy; in this respect,
it should be noted that
no major approximation is made after Eq.~\eqref{App:Eq:scatt1}. The deviations
from that formula are all contained in Eq.~\eqref{App:Eq:resmagscat}.

Although different from the usual resonant function, the energy dependent factor
in Eq.~\eqref{App:Eq:resmagscat} still has for its imaginary part,
a Lorentzian-like shape near its peak. A useful quantity for comparison with
theory and/or experiment is the integral, $I_c$ (units of $r_e\times eV$), of this peak for a resonance
or a group of resonances.
At some distance
from the peak, neglecting $\Gamma_c$, the resonant magnetic scattering length may be
written as
\begin{equation}
f_{res,\,mag,\,c}(\hbar\omega) r_e=
\frac{2 I_c}{\pi}
\frac{(E_a - E_c)^2}{\hbar\omega((E_a - E_c)^2 - \hbar\omega^2)}
\label{App:Eq:normmagscat}
\end{equation}

Consider now the contributions from the Tb $M_{4,5}$ edges.
Calculated values for $I_c$ are in the range 110 ({\it 37, 38}) to
300 ({\it 39}) in units of $r_e eV$ for a fully saturated Tb 4f moment. Thus from
Eq.~\eqref{App:Eq:normmagscat} we find at 6.2$\,keV$ that the resonant magnetic scattering contribution from the Tb $M_{4,5}$
edges to be in the range 0.0005$\,r_e$ to 0.0013$\,r_e$. For our experimental,
and accounting for form factors, the non-resonant magnetic scattering is of
the order of 0.003$\,r_e$ per $\mu_B$, that is 0.027 $r_e$ for a fully saturated Tb moment.
The relative resonant contribution of Tb is therefore between 2\% and 5\% of the non-resonant
contribution. For the Mn L edge, we are unaware of any relevant reliable
calculations or experimental work. Instead we refer to measurements on
the neighbouring  element Fe ({\it 40}).  For metallic
Fe a peak circular dichroic signal of  $\beta=0.005$ was found at the $L_3$ edge with a
peak FWHM=2 $eV$, corresponding to a value of $I_c$ of approximately
$100$ $r_eeV$. (Note: this value is an extreme upper limit
on the value of $I_c$. The total amplitude is expected to be the sum of the contributions from the L$_2$ and L$_3$ edges which is proportional
to the orbital moment.
Since the Mn orbital moment is largely quenched, the contribution from the L edges is expected to be small.)
Then, applying Eq.~\eqref{App:Eq:normmagscat} with
$E_c - E_a = 0.650\, keV$ for Mn $L$ edge, we obtain an amplitude of 0.0001$\,r_e$ at
6.2$\, keV$, that is 1\% of the non-resonant magnetic scattering of an atom
bearing 3$\,\mu_B$.

These considerations lead us to the inexorable conclusion that,
for the energies used in our experiment,
the main contribution to the resonant magnetic scattering amplitude comes from the
Tb $M_{4,5}$ edges, and that this contribution constitutes at most 5\% of the total
magnetic scattering.

Second, we can assess the likely contribution of resonant scattering processes to the
total scattering amplitude relevant for our experiments by considering the available literature.
Of the studies that have been performed to date to investigate non-resonant X-ray magnetic scattering
none, as far as we are aware, have reported issues related to unexpected contamination from
resonant processes. Relevant examples include a recent study of HoMn$_2$O$_5$ ({\it 41}), where the non-resonant
magnetic scattering was measured at  6.4 keV and found to be in perfect agreement with a complex model
of the magnetic structure  deduced from neutron scattering; measurement of the ratio of the orbital to spin magnetization
densities in metallic Holmium at 10.4 keV where good agreement was found with theory ({\it 42});
related experiments on ferromagnetic HoFe$_2$ ({\it 43}); and GdCo$_2$Ge$_2$, where both resonant
and non-resonant magnetic scattering was investigated, again with good internal consistency ({\it 44}).
From the good agreement between theory and experiment in these and other studies
we estimate that the resonant contribution is at most 5-10\% of the non-resonant scattering for the energies relevant
in our experiment.

\begin{figure}
\centering
\includegraphics[width=.485\textwidth]{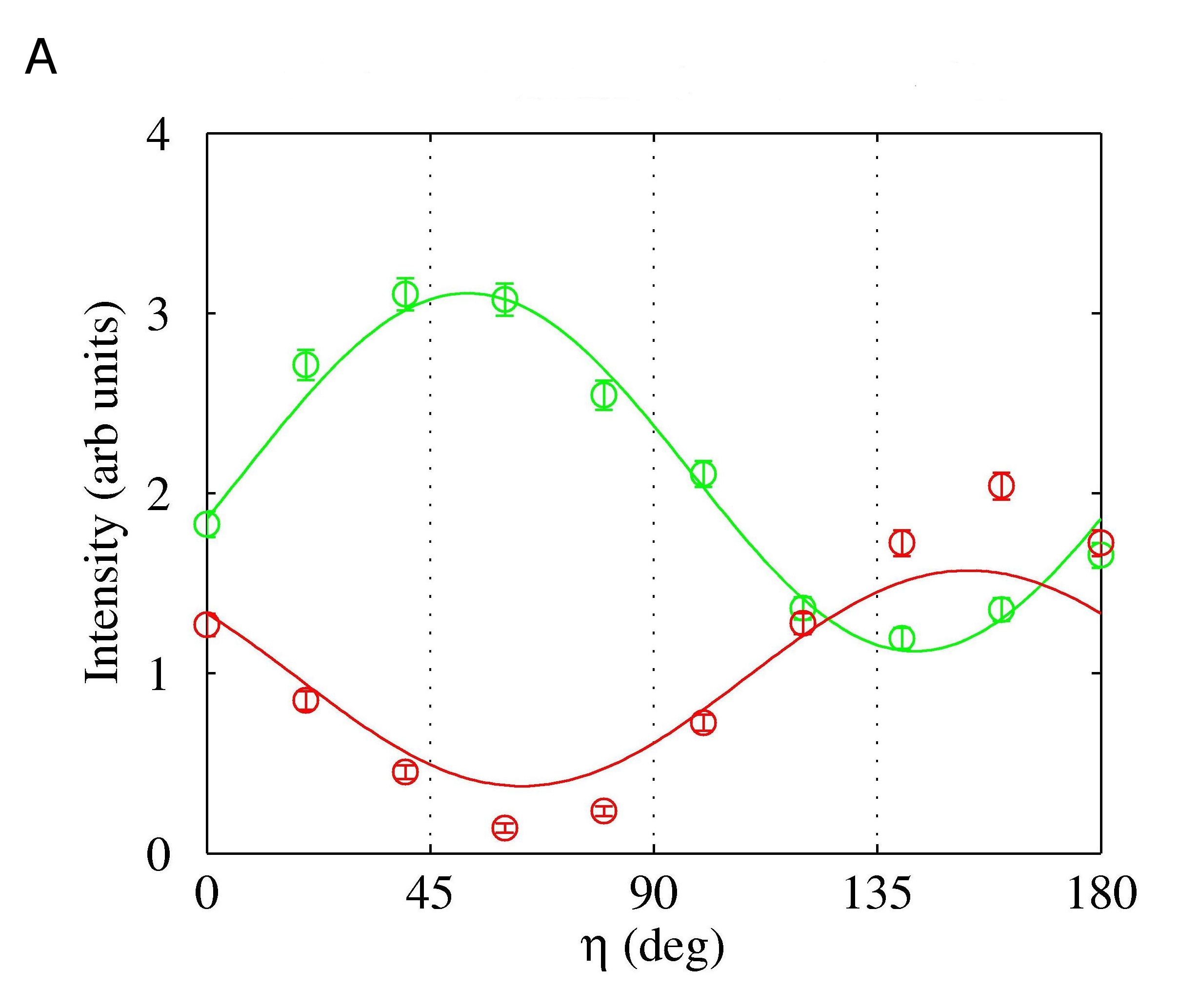}
\includegraphics[width=.485\textwidth]{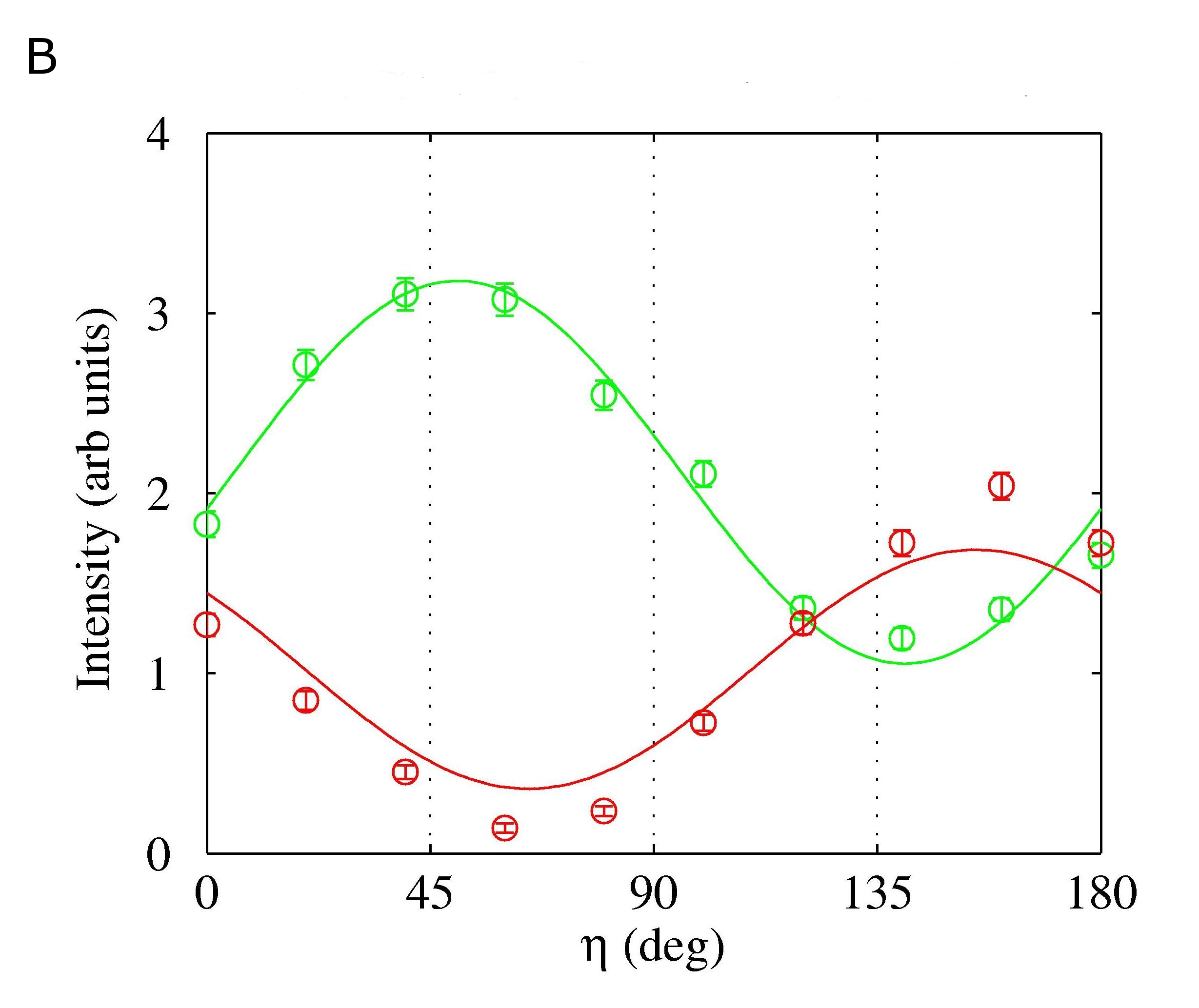}
\caption{The variation with analyzer rotation angle $\eta$ of the x-ray scattering at (4 $\tau$ -1) in TbMnO$_3$ at $T=15$~K, $E=6.85$~keV, in an applied magnetic field of $H=2$~T after a positive electric field cooling, compared with different scattering models: (A) non-resonant magnetic scattering plus ($-0.54(7) - 1.07(6)i$) times charge scattering; (B) non-resonant magnetic scattering plus 0.15 times resonant scattering, and ($-0.58(7) - 1.15(6)i$) times charge scattering.}\label{fig_H2_NRvsRNR}
\end{figure}

Third, we have reanalysed our data to determine whether they are compatible with a significant resonant contribution,
and examined the question of to what extent any such resonant contribution might affect the values of the ionic displacements deduced in
our study. Our starting point was to consider the zero magnetic field data on TbMnO$_3$
presented in reference ({\it 5}). Assuming the magnetic model of Fabrizi ({\it 5}), we found that a reasonable fit to the data was maintained including up to a resonant contribution 15\% of the non-resonant one. The data were also well reproduced using the magnetic model of Kenzelmann ({\it 3}) derived from neutron scattering experiments with the inclusion of a 50\% resonant contribution. (This is not wholly unexpected since, to a reasonable approximation, the resonant term will enter into the Jones matrix in the same way as the non-resonant term arising from the Tb orbital moment.) However, such a model is discounted since (a) it would require a resonant contribution an order of magnitude higher than our theoretical estimate, and (b) it ignores the Tb $b$ moment component which is allowed by group theory. Returning to the model of Fabrizi ({\it 5}) plus a 15\% resonant contribution, the influence of that contribution was assessed by refitting the data in applied magnetic fields. This revealed a negligible effect on the values obtained for the charge scattering, as demonstrated in Figure~\ref{fig_H2_NRvsRNR}, which makes a comparison of fitting the data with and without a 15\% resonant contribution, and indicates that the complex charge scattering amplitude deduced from the two fits is, within error, the same.

Therefore, in conclusion, we find that Eq. 1 is valid to a very good approximation in our experiments, and that any
resonant magnetic contributions that may be present have negligible effect on the derived values of the
ionic displacements.

\clearpage

\clearpage
\begin{table}
\centering
\begin{tabular}{c|cccc}
 & 1 & $2_y$ & $m_{xy}$ & $m_{yz}$\\
\hline
$\Gamma_1$& 1 & 1 & 1 & 1\\
$\Gamma_2$& 1 & 1 & -1 & -1\\
$\Gamma_3$& 1 & -1 & 1 & -1\\
$\Gamma_4$& 1 & -1 & -1 & 1\\
\end{tabular}
\caption{Irreducible representations of the group $G_\mathbf{k}$ for the incommensurate magnetic structure with $\mathbf{k}=(0,\tau,0)$ ({\it 3}).}\label{tab1}
\end{table}

\begin{table}
\centering
\begin{tabular}{|c|c|c|c|c|}
\hline
mode & a & b & c & sites\\
\hline
$\Delta_{\alpha+}$ &                     & $\Gamma_1$ $R$ ext. & $\Gamma_4$ $I$ ext. & \\
$\Delta_{\alpha-}$ &                     & $\Gamma_4$ $R$ ext. & $\Gamma_1$ $I$ ext. & \\
$\Delta_{\beta+}$ &                     & $\Gamma_4$ $I$ vis. & $\Gamma_1$ $R$ vis. & O$_2$ Mn \\
$\Delta_{\beta-}$ &                     & $\Gamma_1$ $I$ vis. & $\Gamma_4$ $R$ vis. & O$_2$ O$_1$ Tb\\
$\Delta_{\gamma+}$ & $\Gamma_1$ $R$ vis. &                     &                     & O$_2$ O$_1$ Tb\\
$\Delta_{\gamma-}$ & $\Gamma_4$ $R$ vis. &                     &                     & O$_2$\\
$\Delta_{\delta+}$ & $\Gamma_4$ $I$ ext. &                     &                     & \\
$\Delta_{\delta-}$ & $\Gamma_1$ $I$ ext. &                     &                     & \\
\hline
\end{tabular}
\caption{The status of all displacive modes. For each component $a,b,c$ of the modes $\Delta$ the irrep to which it belongs, the phase relative to the magnetic component $m_b^{Mn}$ -$R$eal or $I$maginary, and whether they are visible or extincted for the A-type peak in the experiment, are given. Half of the components are absent, since they do not belong to $\Gamma_1$ or $\Gamma_4$. The sites column indicates which sites have the visible modes (shown in fig.~\ref{fig_modes}) in their structure factor.}\label{tab2}
\end{table}

\end{document}